\documentclass[reprint,onecolumn,amsmath,amssymb,aps,superscriptaddress]{revtex4-1}
\usepackage{amsmath}
\usepackage{graphicx}
\usepackage{dcolumn}
\usepackage{bm}
\usepackage{color}
\usepackage{multirow}
\usepackage{epstopdf}
\usepackage{float}
\usepackage{ulem}
\usepackage[export]{adjustbox}
\usepackage{wrapfig}
\usepackage{comment}
\usepackage{natbib}
\usepackage{adjustbox}
\usepackage{soul}
\usepackage{appendix}
\usepackage{hyperref}

\preprint{APS/123-QED}
\begin{document}

\title{Phonon-mediated thermal transport at exponentially mass-graded interfaces: A computational study}

\author{Rouzbeh Rastgarkafshgarkolaei}
\email{rr3ay@virginia.edu}
\affiliation{Mechanical and Aerospace Engineering, University of Virginia, Charlottesville, VA}%

\author{Jingjie Zhang}
\affiliation{Electrical and Computer Engineering, University of Virginia, Charlottesville, VA}%

\author{Carlos A. Polanco}
\affiliation{Materials Science and Technology Division, Oak Ridge National Laboratory, Oak Ridge, Tennessee 37831, USA}%

\author{Nam Q. Le}
\affiliation{National Research Council Research Associateship Programs, U.S. Naval Research Laboratory, Washington, DC}%

\author{Avik W. Ghosh}
\affiliation{Electrical and Computer Engineering, University of Virginia, Charlottesville, VA}%

\author{Pamela M. Norris}
\email{pamela@virginia.edu}
\affiliation{Mechanical and Aerospace Engineering, University of Virginia, Charlottesville, VA}%

\date{\today}

\begin{abstract}
 We numerically investigate thermal transport at solid-solid interfaces with graded intermediate layers whose masses vary exponentially from one side to the other. Using Non-Equilibrium Green's Function (NEGF) and Non-Equilibrium Molecular Dynamics (NEMD) simulations, we show that an exponentially mass-graded junction with a finite thickness can result in 68\% of enhancement in thermal conductance larger compared to a single bridging layer (29\%) and a linear mass-graded junction (64\%) of similar thickness. We examine how the thermal conductance at such interfaces is influenced by geometric qualities and strength of anharmonicity. For geometric properties, we tested the effects from number of layers and the junction thickness. In the absence of anharmonicity, increasing the number of layers results in better elastic phonon transmission at each individual boundary, countered by the decrease of available conducting channels. Consequently, in the harmonic regime, conductance initially increases with number of layers due to better bridging, but quickly saturates. The presence of slight anharmonic effects (at ultra-low temperature \textbf{T} = 2 K) turns the saturation into a monotonically increasing trend. Anharmonic effects can facilitate interfacial thermal transport through the thermalization of phonons. At high temperature, however, the role of anharmonicity as a facilitator of interfacial thermal transport reverses. Strong anharmonicity introduces significant intrinsic resistance, overruling the enhancement in thermal conduction at the boundaries. Our analysis shows that in our model Lennard-Jones system, the influence of a mass-graded junction on thermal conductance is dominated by the phonon thermalization through anharmonic effects, while elastic phonon transmission plays a secondary role. It follows that at a particular temperature, there exists a corresponding junction thickness at which thermal conductance is maximized.
\end{abstract}
\pacs{Valid PACS appear here}
\maketitle
\[\]
%

\section{Introduction}
New challenges for  thermal management of semiconductor devices have arisen due to the miniaturization of present day electronics to the nanoscale\cite{Pop2010}. For such devices, thermal resistance at material interfaces limits heat dissipation, increases their operating temperature, and ultimately impacts their performance and reliability~\cite{Riedel2009}. The heat dissipation problem of semiconductor devices can be mitigated using high thermal conductivity materials, like diamond, as heat spreaders. However, this approach is limited by the thermal resistance at material interfaces arising inside and in between devices, as well as their connections to external bias and contact pads~\cite{Pomeroy2014,Sun2016,Zhou2017,Cho2017}. Thus, thermal resistance at interfaces is a critical bottleneck for thermal management of semiconductor devices and concerted efforts are now focused on reducing this resistance~\cite{Cahill2003,Hopkins2013,Cahill2014}.

The thermal conduction at an interface can be enhanced by varying interfacial properties such as roughness\cite{Merabia2014}, atomic composition and bonding \cite{Monachon2016, Zhang2018, Rastgar2016, Aryana2018}. By strengthening the bonds at a junction for example, phonon transmission across the interface can be made to increase, along with the corresponding thermal conductance~\cite{Losego2012,Hohensee2015,Schmidt2010,Saltonstall2013,Duda2013,Jeong2016}. The interfacial geometry can also influence thermal transport across material interfaces by providing larger effective interfacial contact area~\cite{Lee2016a}. Even interatomic mixing at the interface can result in an increase in conductance due to the introduction of new transport channels across the interface \cite{English2012b,Tian2012,Polanco2015}. Moreover, it has been shown that anharmonic interactions significantly facilitate heat transfer across solid-solid interfaces~\cite{saaskilahti2014}.

It has been proposed to enhance interfacial conductance by insertion of a thin ($\sim$nm) intermediate layer at the interface~\cite{Liang2011,English2012b,Duda2013,Smoyer2015,Polanco2016,Polanco2017,lee2017a} similar to applications of anti-reflective (AR) coatings in photonics~\cite{Raut2011}. In the harmonic limit, the layer increases the elastic transmission of phonon modes as well as the overlap of phonon density of states (PDOS) of the materials at the newly formed interfaces and acts as a ``phonon bridge" \cite{English2012b}. Previously, we showed that anharmonic processes play a key role in the enhancement of thermal conduction in these systems in comparison with the purely harmonic limit~\cite{Polanco2017}. At each material junction, those processes help phonons thermalize to frequencies with higher transmissions and thus they can increase the thermal conductance~\cite{Le2017}. Moreover, anharmonic processes decouple the two material interfaces abutting the thin layer and thus the system resistance can be represented as the sum of the boundary resistances plus a junction resistance~\cite{Polanco2017}. This generates an optimum condition: since each interfacial resistance depends on the ratio of the acoustic impedances on each side, the maximum thermal conductance happens when the atomic mass (i.e., impedance for constant bond stiffness) of the layer is the geometric mean of the contact masses\cite{Polanco2017}, which we refer to as the ``geometric mean rule" throughout the rest of this article.

In this work, we refer to the {\it{additive}} transport regime, in which interfaces become decoupled and we observe that thermal resistances are additive due to incoherence~\cite{Datta2005}. Commonly, it is thought that the overall resistance at a bridged interface equals the sum of interfacial and intrinsic resistances only when transport is diffusive, i.e. when phonons mean free path (MFP) becomes smaller than the layer thickness. However, we have shown previously~\cite{Polanco2017} that at bridged interfaces, layer thicknesses much smaller ($\sim$ nm) than the bulk phonon MFP are sufficient for the thermal resistance to be treated as the sum of the resistances. Based on our definition, the diffusive limit is the extreme limit for the additive regime in terms of increasing phonon scattering.

\begin{figure}[h!]
	\centering
	\includegraphics[width=80mm]{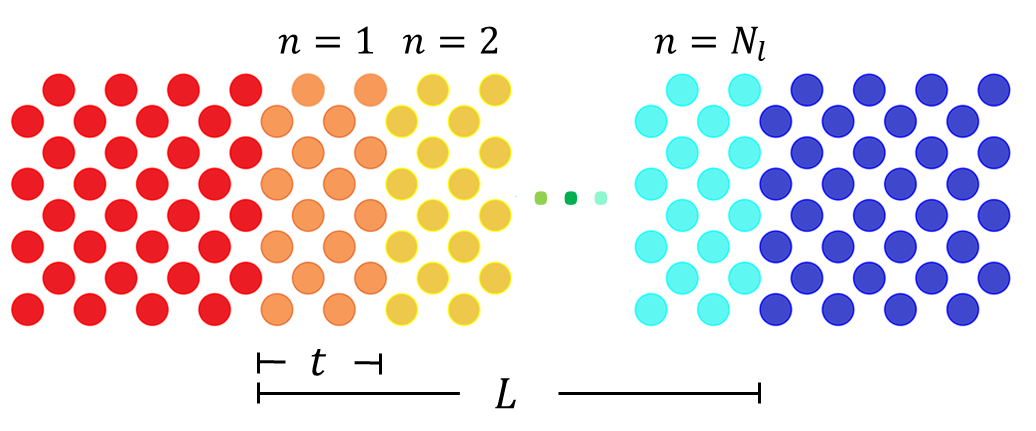}
	\caption{Schematic of a mass-graded interface with $N_l$ layers. In this case, each layer has a thickness of 2 unit cells ($t = 2$ u.c.) and the thickness of the junction is $L = t \times N_l$ u.c.}
	\label{fig1}
\end{figure}

Building on our previous work with a monolithic bridging interface~\cite{Le2017,Polanco2017}, in this paper we explore the enhancement of thermal conductance of a {\it{mass-graded}} interface or an interface with several intermediate thin layers  (Fig.~\ref{fig1}) analogous to the design of refractive index-graded AR coatings~\cite{Clapham1973, Bernhard1967}. The atomic mass of each layer $(m_n)$ is chosen based on the geometric mean rule relative to its neighboring layers, which corresponds to an exponential change of the atomic masses from the left contact mass $(m_l)$ to the right contact mass $(m_r)$ described by:
\begin{equation}
m_n = m_l.e^{\zeta n},
\label{eqnexp}
\end{equation}
with $\zeta = \ln{(m_r/m_l)}/{(N_l+1)}$. We show that this choice of masses leads to a conductance enhancement about two times larger than the best enhancement obtained with a single bridging layer studied by Polanco {\it{et al.}}~\cite{Polanco2017} (see Fig.~\ref{fig8} for $L=6$ u.c.). Moreover, we demonstrate larger enhancement compared to a previously proposed linearly mass-graded interface\cite{Zhou2016} (Sec.~\ref{seclinvsexp}). Our results examine the influence of the number of layers $N_l$ and the thickness of the layers $t$ (Fig.~\ref{fig1}) on the thermal conduction across mass-graded interfaces.

Besides studying the influence of geometric parameters, we also explore the effect of varying the strength of anharmonicity on the conductance. Non-equilibrium Green's function (NEGF) and Non-Equilibrium Molecular Dynamics (NEMD) simulations are used to compare interfacial thermal transport without anharmonic processes at \textbf{T} = 0 K (Sec.~\ref{sechar}), with weak anharmonicity at low temperature of 2 K (Sec.~\ref{secwanhar}), and with strong anharmonicity at medium temperature of 30 K (Sec.~\ref{secsanhar}). We find that the strength of anharmonicity determines how different geometric properties of the mass-graded junction influence the conductance. In the limit of weak anharmonicity, increasing the junction thickness facilitates thermal transport by phonon thermalization. In the limit of strong anharmonicity, however, increasing the layer thickness over the optimum thickness increases phonon back scattering and suppresses thermal transport. Our results suggest that in our model Lennard-Jones system, anharmonic effects contribute mostly to the enhancement in the conductance associated with insertion of a mass-graded junction while contributions from elastic phonon transmission come secondary.
%
%
%
%
%
%
\section{Methodology}

The focus of our study is mass-graded interfaces (Fig.~\ref{fig1}), with the atomic mass of each intermediate layer varying exponentially from the left to the right contact according to Eq.~\ref{eqnexp}. All atomic interactions in the system are dictated by the same Lennard-Jones (LJ) potential (see Appendix A). The system has a single atom per primitive unit cell and a face-centered cubic crystal structure. Thermal conductance ($G$) across the mass-graded interface is defined as the ratio between the heat flux ($q$) crossing the interface and the temperature drop across the entire junction ($\Delta$\textbf{T}):
\begin{equation}
    G = \frac{q}{\Delta \textbf{T}}.
    \label{equG}
\end{equation}
To calculate $G$ using Eq.~\ref{equG} within NEMD, we prescribe a constant temperature difference over the simulation box. Upon reaching a steady state temperature profile, we fit the temperature data at the contacts with linear profiles, which are extrapolated to the external edges of the first and last intermediate layers to define $\Delta \textbf{T}$. Heat flux $q$ is calculated by monitoring the cumulative energy added/subtracted to the hot/cold Langevin baths. We calculate thermal conductance at $\textbf{T} = 2$ K and $\textbf{T} = 30$ K, which are 0.7\% and 10\% of the melting temperature, to explore the phonon transport in the limit of weak and strong phonon-phonon interactions.
The conductance in the limit of zero phonon-phonon interactions is calculated using harmonic NEGF~\cite{Jeong2012,ghosh2016}. To compare these simulations with the NEMD results, we take the classical limit of the Bose-Einstein distribution ($\hbar\omega_{cut}\ll k_B \textbf{T}$) and compute the conductance as~\cite{Polanco2017}:
\begin{equation}
\label{equGhl}
G=\frac{k_B}{2\pi A}\int_0^\infty d\omega MT,
\end{equation}
where $\hbar\omega_{cut}$ is the maximum phonon energy,  $k_B$ is the Boltzmann constant, $A$ is the cross-sectional area, $M$ is the number of modes contributing to transport, and $T$ is the average transmission per mode. $MT$ is calculated using the NEGF formalism as $MT= \text{Tr}[\Gamma_lG^r\Gamma_rG^{r\dagger}]$, where $G^r$ is the retarded Green's function describing the vibrational dynamics of the interface and $\Gamma_{l,r}$ are the broadening matrices describing how the modes in the contacts interact with the intermediate layers~\cite{Datta2005,Mingo2003,Wang2008}. Equation~\ref{equGhl} shows that $G$ is a summation of modes times transmission over the entire frequency spectra, meaning that $M$ and $T$ are the two components determining the conductance. Further details of all calculations are provided in Appendix A.

\section{Harmonic limit} 
\label{sechar}

The analysis of our mass-graded interfaces in the harmonic limit is simplified using the system symmetry. Since all the material boundaries are perfectly abrupt, the potential energy is translationally invariant in the transverse direction, parallel to the boundaries. Thus, the force in that direction is zero and only phonons that conserve their transverse momentum or wavevector ($k_\perp$) can contribute to thermal transport. We define the number of combinations of phonons that conserve momentum along the system as the number of conserving channels $M_c$ and count them using\cite{Polanco2017}
\begin{equation}
    M_c(\omega)=\sum_{k_\perp}\min_\alpha M_\alpha(\omega,k_\perp),
    \label{eqnMc}
\end{equation}
with $\alpha$ varying over the contacts and intermediate layers. $M_\alpha$ is the number of propagating modes in material $\alpha$, which can be obtained by calculating $MT$ from NEGF for each bulk material. In that case, the transmission for each mode is unity and thus $MT=M$. Since the conserving modes are the only ones that contribute to transport, we define an average transmission over those modes as $T_c(\omega)=MT(\omega)/M_c(\omega)$. Replacing $MT$ in Eq.~\ref{equGhl} by $M_cT_c$ allows us to separate $G$ into a phase space of available transport channels, $M_c$, and its average phonon transmission, $T_c$. 

\begin{figure}[ht]
	\centering
	\includegraphics[width=80mm]{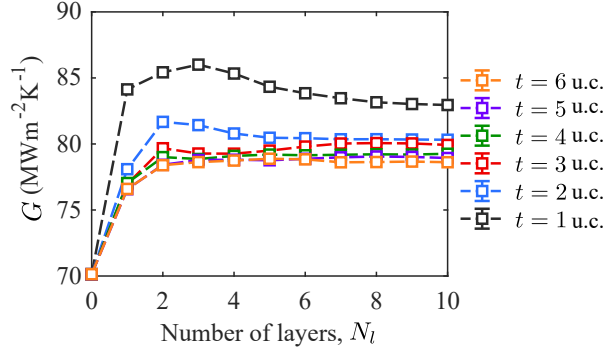}
	\caption{$G$ vs. $N_l$ in the harmonic limit. $G$ quickly saturates as $N_l$ increases.}
	\label{fig2}
\end{figure}

\begin{figure}[ht]
	\centering
	\includegraphics[width=80mm]{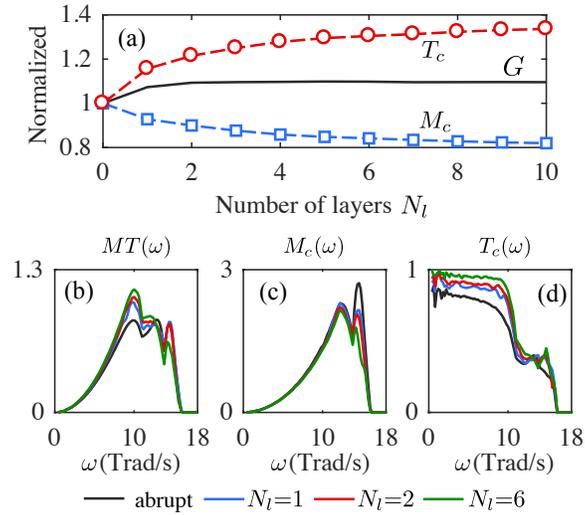}
	\caption{(a) Normalized values of $G$, $M_c$, and $T_c$ with respect to the abrupt interface vs. $N_l$. $M_c=\int_0^\infty{M_c(\omega)}d\omega$ and $T_c=\int_0^\infty{T_c(\omega)}d\omega$. $T_c$ increases while $M_c$ decreases with $N_l$, leading to the saturation of $G$. NEGF results of (b) number of modes times transmission MT($\omega$), (c) number of available modes $M_c(\omega)$ and (d) average transmission $T_c(\omega) =\frac{MT(\omega)}{M_c(\omega)}$ when $N_l$ is 0 (abrupt), 1, 2 and 6. All simulations are performed for $t=6$ u.c.}
	\label{fig3}
\end{figure}

As the  number of intermediate layers, $N_l$, of a mass-graded interface increases, the harmonic conductance initially increases  but saturates after $N_l>5$ (Fig.~\ref{fig2}). This trend is due to the interplay (see Eq.~\ref{equGhl}) between increasing  transmission $T_c$ (Fig.~\ref{fig3}(a) and (d)) but decreasing number of transport channels $M_c$ (Fig.~\ref{fig3}(a) and (c)). The gain in $T_c$ is due to the decrease in thermal impedance (acoustic impedance in linear dispersion regime) mismatch between adjacent layers~\cite{Cahill2003,Polanco2013}. This gain happens mostly below 10 Trad/s (Fig.~\ref{fig3}(d)) and is responsible for the increase of $MT(\omega)$ over the same frequency range (Fig.~\ref{fig3}(b)) since $M_c$ does not change much in that range. Note that the cut-off frequency for the lowest acoustic branch is 10.98 Trad/s, which seems to suggest that decreasing the mass mismatch helps phonon transmission for states with similar polarization (Fig.~\ref{fig1S}). The monotonic decrease of $M_c$ follows from Eq.~\ref{eqnMc} as adding more intermediate layers can only decrease the minimum of modes at each $k_\perp$ and $\omega$. The interplay between $M_c$ and $T_c$ yields a modest conductance enhancement in the saturated regions ($N_l > 5$ in Fig.~\ref{fig2}), between 11\% and 17\%.

The saturation of $G$ follows from a combined saturation of $M_c$ and $T_c$. $M_c(\omega, k_\perp)$ is obtained taking the minimum of modes (Eq.~\ref{eqnMc}) over a set of materials with the same force constants and crystal structure, but with masses varying exponentially from one contact to another. Thus the dispersions and $M_\alpha(\omega,k_\perp)$ for those materials change gradually according to the mass. As $N_l$ increases, the interval of this function is sampled more finely by the set of $M_\alpha(\omega,k_\perp)$, and thus $M_c$ saturates to the lower bound. The transmission enhancement also saturates as it approaches its maximum value, unity (Fig. \ref{fig3}d). 

The conductance of a mass-graded junction does not only depend on the number of layers, it also depends on the thickness of each layer $t$ (Fig.~\ref{fig2}). Thin layers yield larger conductance, but this enhancement disappears at about $t=3$ u.c. We attribute the sharp increase in $G$ when the layer thickness is ultra-thin to phonon tunneling. For very thin layers (in our case, 2--3 conventional unit cells), phonons can tunnel even when the middle layers do not have propagating modes at a particular $\omega$ and momentum $k_\perp$ but the adjacent materials do. The transport of those extra phonons across the system enhances the overall conductance. This phenomenon was previously observed by English {\it{et al.}}~\cite{English2012b} and Liang and Tsai~\cite{Liang2011} and they related it to the resulting sharp and narrow density of states associated with the thin film which can influence the elastic vs. inelastic thermal transport at the boundaries.

\section{Weakly anharmonic limit} 
\label{secwanhar}

\begin{figure}[ht]
	\centering
	\includegraphics[width=80mm]{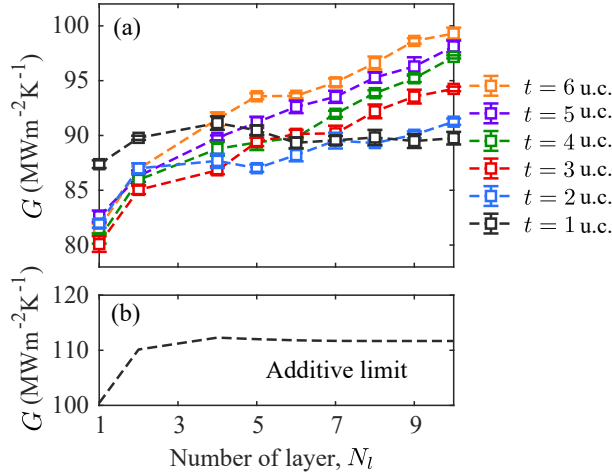}
	\caption{$G$ vs. $N_l$ in the presence of anharmonicity at \textbf{T} = 2 K when the layer thicknesses vary from 1 u.c. to 6 u.c. (a) NEMD results. $G$ increases almost linearly with $N_l$. Furthermore thicker layers yield larger $G$. b) additive limit (Eq. \ref{eqnresistance}). $G$ increases with $N_l$ and quickly saturates.}
	\label{fig4}
\end{figure}

Surprisingly, at low temperature when anharmonicity is weak, the trend of $G$ vs. $N_l$ from our NEMD simulations (Fig.~\ref{fig4}(a)) differs from that obtained in the harmonic limit by NEGF. We were expecting similar trends because at low temperature ($\textbf{T} = 2$ K, which is about $1\%$ of the melting temperature), atomic displacements in our NEMD simulation are small and thermal transport should be mostly harmonic. Nevertheless, this expectation seems to hold only for systems with $t=1$ u.c., where we see a peak followed by a saturation (Fig.~\ref{fig2} and \ref{fig4}(a)). We have verified that the observed trends do not result from size effects on the simulation domains (see Appendix A).

The increasing trend of $G$ vs $N_l$ in our ultra-low temperature NEMD simulations (Fig.~\ref{fig4}) is not dictated by additive phonon transport either. In the additive limit, the conductance of the system, $G_{dl}$, can be defined as the inverse of the sum of resistances:
\begin{equation}
    1/G_{dl} = \sum_{i=1}^{N_l} 1/G_{blk,i} + \sum_{j=1}^{N_l+1}1/G_{int,j},
\label{eqnresistance}
\end{equation}
where $1/G_{blk,i}=t/\kappa_i$ is the resistance intrinsic to the $i^{th}$ intermediate layer, $\kappa_i$ is the intrinsic thermal conductivity of material $i$ and $1/G_{int,j}$ is the interfacial resistance for the $j^{th}$ boundary. We neglect $1/G_{blk,i}$ in our analysis since it is significantly less than $1/G_{int,j}$ at $\textbf{T} = 2$ K. For instance for a mass-graded interface with $t=6$ u.c. and $N_l=5$, the temperature drop at the interfaces is 93\% of the total drop between the contacts (Fig. ~\ref{fig2S}). Figure~\ref{fig4}(b) shows the trend of $G$ vs.\ $N_l$ in the additive limit with each $G_{int,j}$ calculated on a single, independent boundary using NEGF and neglecting $1/G_{blk,i}$. $G$ initially increases as neighboring layers become more similar and then saturates. The saturation is not seen in NEMD results and thus we conclude that the monotonic increase of conductance at very low temperatures results from neither purely harmonic nor additive transport.

The increasing trend in Fig.~\ref{fig4} hints at the important role played by phonon-phonon interaction in enhancing the conductance of mass-graded interfaces. Conductance seems to increase linearly with $N_l$ and the slope increases with $t$. Larger $N_l$ and $t$ values result in a thicker total junction length, $L$, which allows more phonon-phonon scattering in this region. Given the conductance increases as phonon-phonon scattering increases, we hypothesize that scattering promotes thermalization that helps high frequency phonons with lower chance of transmission jump to modes with lower frequencies and higher transmission. This behavior is similar to the linear increase of interfacial thermal conductance with temperature, in which stronger anharmonicity contributes to better thermalization in the neighborhood of the interface{\cite{Le2017,Polanco2017}}.

\begin{figure}[ht]
	\centering
	\includegraphics[width=80mm]{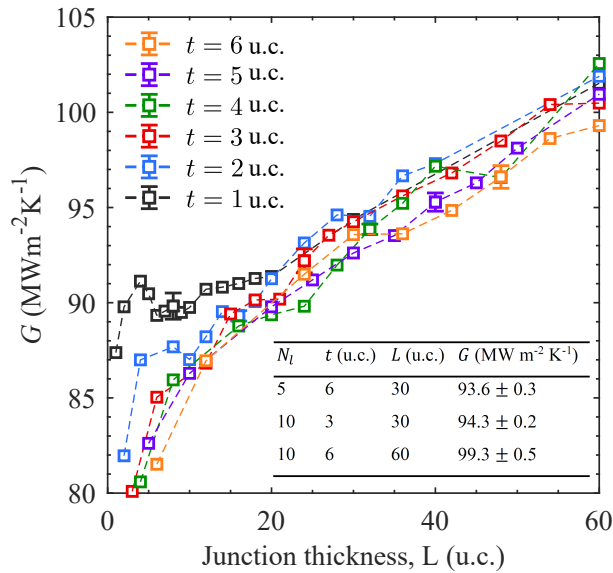}
	\caption{Interfacial thermal conductance values from NEMD simulations at \textbf{T} = 2 K for different junction thicknesses. Each color represents a different sub-layer thickness.  Note that total thickness $L$ = $N_l$ $\times$ $t$. Sample error bars are shown at L = 8, 16, 24, 32, 40, and 48 u.c.\ (inset) conductance values are shown for three cases of varying number of layers, layer's thickness and junction thickness.}
	\label{fig7}
\end{figure}

The contributions to the enhancement of $G$ from both anharmonicity and elastic phonon transmission are further analyzed in Fig.~\ref{fig7}. Conductance increases with $L$ with a similar slope when $t>$1, suggesting that anharmonicity constitutes the major contribution in the enhancement. This idea is further supported by comparing the enhancement from varying $N_l$ while fixing $L$ (i.e. varying phonon transmission at a fixed strength of anharmonicity) with the results from fixed $N_l$ while varying $L$ (varying the strength of anharmonicity with fixed phonon transmission). Figure~\ref{fig7} suggests that the contribution from the latter is larger than the former. To make this argument quantitative, we turn to the conductance values shown in the inset of Fig.~\ref{fig7}. At a fixed $L = 30$ u.c., doubling $N_l$ results in less than 1\% enhancement in $G$, whereas at a fixed $N_l$, increasing $L$ from 30 to 60 u.c. results in more than 5\% improvement in conductance.  When fixing $L$, the enhancement would solely be due to increases in phonon transmission at the boundaries; however, this enhancement is very small without the presence of anharmonicity. Bridging layers not only introduce better matching at each boundary, but also provide phonons with opportunity for thermalization, providing thereby a larger contribution to the overall enhancement. 

\begin{figure}[ht]
	\centering
	\includegraphics[width=80mm]{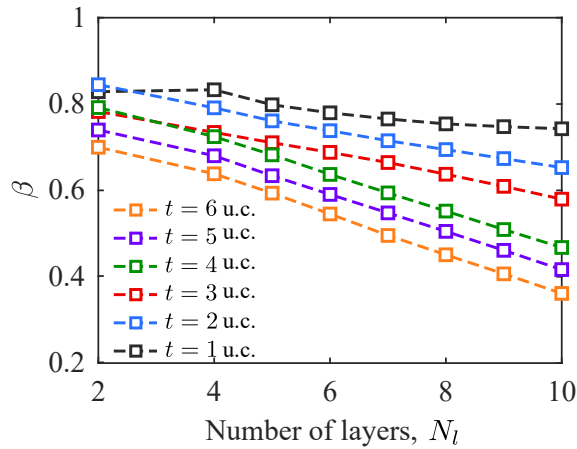}
	\caption{The contribution from ballistic phonon transport $\beta$ vs $N_l$ when the layer thickness varies from 1 unit cell to 6 unit cells. $\beta = 1$ represents purely ballistic transport.}
	\label{fig6}
\end{figure}

The values of $G$ in the weak anharmonic limit (Fig.~\ref{fig4}(a) and \ref{fig7}) are bounded by those in the harmonic limit (lower bound) (Fig.~\ref{fig2}) and those in the additive limit (upper bound) (Fig.~\ref{fig4}(b)). Also, as $N_l$ or $L$ increases, $G$ seems to transition from the harmonic to the additive limit. To quantify the ratio of harmonic vs. additive phonon transport across the junction, we define a quantity $\beta$ such that $G=\beta G_{hl}+(1-\beta)G_{al}$, where the harmonic conductance $G_{hl}$ is obtained from NEGF calculations across multiple layers, while the additive limit conductance $G_{al}$ is obtained from Eq.~\ref{eqnresistance} by adding NEGF calculations at single boundaries (Fig.~\ref{fig6}). As $N_l$ increases, $G$ approaches the additive limit and thus $\beta$ decreases, meaning less phonons can transport across all the interfaces without being scattered by other phonons. This is consistent with our conjecture that the bridging layers facilitate more phonons participating in the thermalization process. 

\section{Strongly anharmonic limit} 
\label{secsanhar}

\begin{figure}[ht]
	\centering
	\includegraphics[width=80mm]{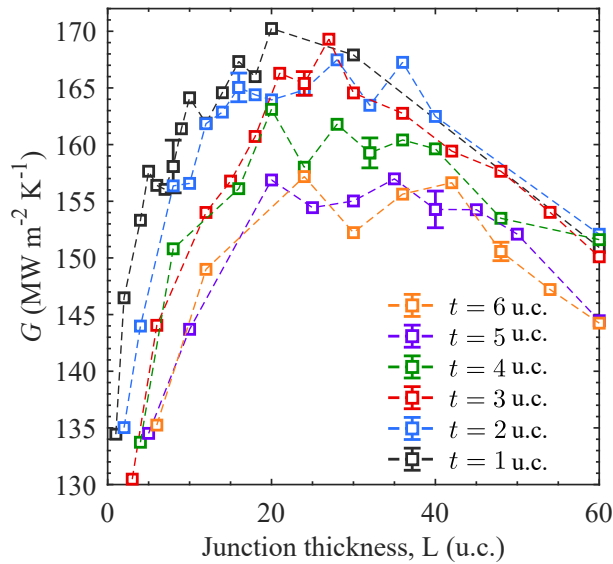}
    \caption{Interfacial thermal conductance from NEMD simulations at \textbf{T} = 30 K for different thicknesses of the junction. Each color represents a different sub-layer thickness. Sample error bars are shown at L = 8, 16, 24, 32, 40, and 48 u.c.}
	\label{fig8}
\end{figure}

Conductance values from NEMD simulations at high temperature ($\textbf{T} = 30$~K), when anharmonicity is strong, are plotted in Fig.~\ref{fig8}. The trend of $G$ vs. $L$ results from the interplay between the interfacial $G_{int,j}$ and intrinsic $G_{blk,i}$ conductances in Eq.~\ref{eqnresistance}. Table~\ref{tab:contribution} explains the opposite influences of these two conductance terms quantitatively. For constant layer thickness $t$, when $N_l$ increases from 2 to 7, the overall conductance $G$ increases because of higher phonon transmission at individual boundaries which increases the interfacial conductances $G_{int,j}$. Moreover, in this regime, extra phonon-phonon scattering provided by larger $L$ enhances the conductance through phonon thermalization, also resulting in higher $G_{int,j}$ values. $G$ then reaches a maximum at junction thicknesses around 20-30 u.c. When $N_l$ increases from 7 to 10, $G$ decreases with the junction thickness because the gain of conductance at the boundaries $G_{int,j}$ is overshadowed by the decrease in the layers' intrinsic conductance $G_{blk,i}$. Consequently, the maximum $G$ is dictated by the interplay between the intrinsic phonon-phonon resistance of the mass-graded junction (1/$G_{blk,i}$) and the interfacial resistance at each individual boundary (1/$G_{int,j}$). 

This interplay is mainly driven by the strength of anharmonic processes in the system. To elaborate, we observed in Fig.~\ref{fig7} that at low temperature when anharmonicity is weak, extra anharmonicity provided by thicker junctions can enhance the transport. On the other hand, in Fig.~\ref{fig8} when anharmonicity is strong, extra anharmonic scattering will be a disadvantage to the overall transport. In both of these scenarios the potential enhancement from higher elastic transmission of phonons at the boundaries is subtle. These observations show that the influence of a bridging layer on thermal conductance is dominated by the phonon thermalization through anharmonic effects.

\begin{table}[h!]
\caption{Contribution from interfacial thermal conductance $G_{int,j}$ at individual boundaries and intrinsic conductance $G_{blk,i}$ values to the overall thermal conductance $G$ at \textbf{T} = 30 K from NEMD for three cases with equal layer thickness $t$ and different number of layers $N_l$.}
\centering
\begin{adjustbox}{width=85mm}
\begin{tabular}{ l l c c c }
 \hline
 \hline
 \\
 $t$ (u.c.) & $N_l$ & $\frac{1}{\sum_{i} 1/G_{blk,i}}$ & $\frac{1}{\sum_{j} 1/G_{int,j}}$ & $G$ (MW m$^{-2}$ K$^{-1}$)\\
 \\
 \hline
 \\
 6 & 2 & 626.5 & 195.4 & 148.9 $\pm$\ 2.0 \\
 \\
 6 & 7 & 299.9 & 325.6 & 156.1 $\pm$\ 1.2 \\ 
 \\
 6 & 10 & 240.1 & 364.8 & 144.8 $\pm$\ 0.8 \\
 \\
 \hline
 \hline
\end{tabular}
\end{adjustbox}
\label{tab:contribution}
\end{table}

\begin{figure}[h!]
	\centering
	\includegraphics[width=76mm]{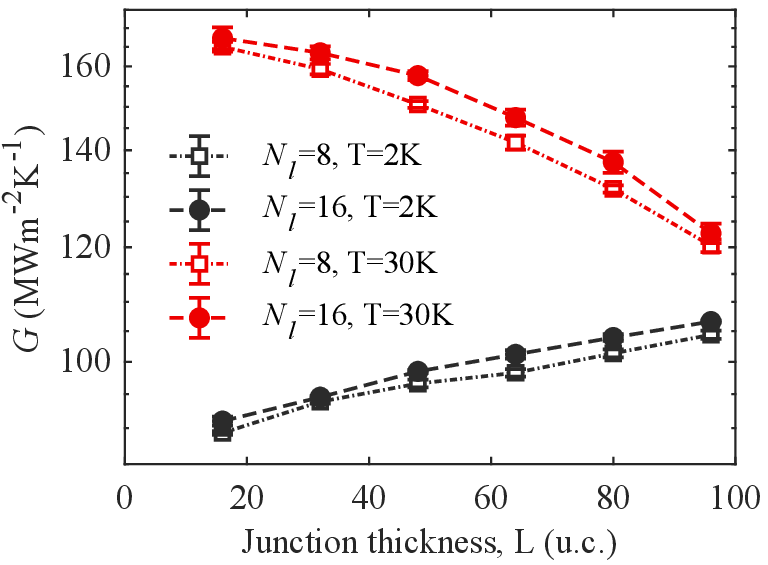}
	\caption{G vs. total thickness of the graded interface, $L$, for fixed number of layers ($N_l$ = 8 vs. 16) at low (\textbf{T} = 2 K) and high (\textbf{T} = 30 K) temperatures.}
	\label{fig5}
\end{figure}

Figure~\ref{fig5} shows the effects of anharmonicity on the thermal conductance of mass graded interfaces, both as a facilitator in the weak anharmonicity limit and a suppressor in the strong anharmonicity limit. We calculated $G$ keeping the number of layers constant at either $N_l = 8$ or $16$, while varying the thickness of each layer $t$ (see Fig.~\ref{fig5}). At low temperature, i.e. weak anharmonicity, $G_{blk,i}$ is negligible and thermal conductance increases with junction thickness. This trend flips at high temperatures, shown in Fig.~\ref{fig5}, where anharmonicity is strong and $G_{blk,i}$ plays a profound role in hindering thermal transport and consequently, $G$ decreases with junction thickness.  

Using the results from this computational work, our goal is to guide experiments to design engineered interfaces with enhanced thermal conductance including bridging interfaces, that can be integrated into devices to achieve superior thermal and electronic performances. It has previously been shown by Wu et al.~\cite{Wu1994} that a symmetric graded layer of GaAs/In$_{x}$Ga$_{1-x}$As/GaAs can enhance the electron mobility. Moreover, experimental fabrication of compositionally graded junctions has been enabled by a metal organic chemical vapor deposition (MOCVD) process and thus necessitates an investigation on the effect of the parameters, which can be adjusted during the fabrication, that can alter the thermal transport~\cite{HoiWong2014}. Change in material composition across the junction is an important factor that can influence the thermal transport at mass-graded interfaces. In the next section, we compare the level of enhancement in $G$ using two different grading schemes: exponential vs.\ linear. 
\section{Exponential vs.\ Linear} \label{seclinvsexp}

A 6 fold increase of thermal conductance was previously reported~\cite{Zhou2016} for linear mass-graded interfaces. In this section, we aim to compare the percentage of enhancement in two systems: grading the mass along the intermediate layer 1- linearly and 2- exponentially. Since the exponential variation of masses results directly from the geometric mean rule, which yields close to maximum conductance at an interface with a single intermediate layer, it is expected that exponentially graded interfaces result in larger enhancement in $G$.

\begin{figure}[ht]
	\centering
	\includegraphics[width=80mm]{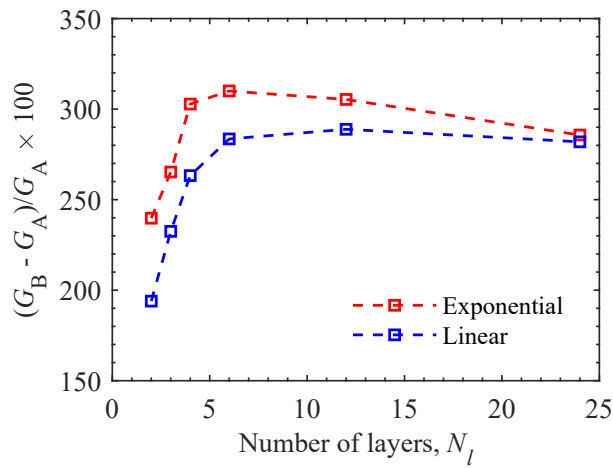}
	\caption{Comparing enhancement in thermal conductance values ($G_B$ and $G_A$ stand for the conductance of the bridged and abrupt interfaces, respectively) between linearly and exponentially mass-graded interfaces, varying the number of layers at the interface, and keeping the total thickness constant. Junction thickness $L$ for all the cases is 12 u.c. System temperature is set to be \textbf{T} = 30 K. ($m_l = 40$ amu , $m_r = 400$ amu)}
	\label{fig10}
\end{figure}

To test the aforementioned hypothesis, we setup the baseline system with mass mismatch of 10 between the two contacts, i.e. $\frac{m_r}{m_l} = 10$, to replicate the work presented by Zhou {\it{et al.}}~\cite{Zhou2016}. We set the junction thickness to be $L=12$ u.c. and the temperature \textbf{T} = 30 K for all the systems, so that the two systems are exactly the same except for the mass of each layer. Our results indicate an extra enhancement in $G$ upon utilizing the exponential mass-graded interface compared to its linear counterpart (Fig. \ref{fig10}). A maximum enhancement of 308\% is attained for the exponential mass-graded interface, compared to the linear grading which gives 289\% ($N_l = 6$ in Fig. \ref{fig10} ). Our results are in line with our previous findings and hypothesis\cite{Polanco2017}, that the enhancement due to a bridging layer can be maximized when the mass of the intermediate layer is close to the geometric mean of the contact masses.

The difference between the conductance values for these two types of interfaces slowly decreases as the number of layers increases. More layers result in smaller mass mismatch at each boundary. Previously\cite{Polanco2017}, we showed that as mass mismatch ($\frac{m_r}{m_l}$) at the interfaces decreases, a wider range of masses around the geometric mean produce a conductance close to the maximum. Thus, the influence of the geometric mean rule reduces for larger numbers of layers and the exponential and linear mass-graded interfaces exhibit a similar enhancement in thermal conduction.

The percentage of enhancement at these mass-graded interfaces strongly depends on the amount of mass mismatch in the systems. Our results indicate that enhancement in $G$ varies from 68\% to 308\% as the mass ratio varies from 3 to 10. It is necessary to develop a framework where the reported percentage of enhancement is independent of the vibrational mismatch. This may be done by taking advantage of the dependency of $G$ from the mass ratio at the boundary and maximum phonon frequency present in the system, as shown in our previous work\cite{Polanco2017}.   

\section{Conclusion}
We have demonstrated that the introduction of an exponentially mass-graded junction can enhance thermal conductance beyond its linear graded counterpart. The enhancement of conductance at such interfaces depends on the number of layers, thickness of the junction and the temperature. In the harmonic limit, increasing the number of layers results in better acoustic impedance matching at the boundaries and higher phonon transmission at those individual interfaces and thus facilitates thermal transport. On the other hand, adding more layers in the junction decreases the number of transport channels, consequently hindering the transport. These opposing actions thus result in increasing the overall conductance initially, and then turns into an asymptotic saturation of thermal conductance when the number of layers is large. We also found that the potential enhancement using a mass-graded junction strongly depends on anharmonicity, which is both influenced by the thickness of the junction and the temperature. Anharmonic processes, however, play opposing roles on the conductance. At low temperature, when anharmonicity is weak, extra anharmonicity provided by the thicker junctions facilitates transport by thermalizing phonons with higher frequencies to the modes with lower frequencies and higher chance of transmission at the boundaries. In the limit of strong anharmonicity, however, the intrinsic resistance of the junction overshadows the gain in conductance at the boundaries, and consequently extra anharmonicity hinders the transport in this regime. Influence of mass-grading on thermal conductance is dominated by the phonon thermalization through anharmonic effects, while elastic transmission of phonon modes across boundaries plays a secondary role. Lastly, we find that the percentage of enhancement strongly depends on the mass mismatch at the interface, varying from 308\% to 68\% as the ratio varies from 10 to 3.\\

\begin{acknowledgments}
R.R and P.M.N. acknowledge the financial support of the Air Force Office of Scientific Research (Grant No. FA9550-14-1-0395). J.Z. and A.W.G acknowledge the support from ``Graduate opportunity (GO!)" program associated with Center for Nanophase Materials Sciences (CNMS) at Oak Ridge National Laboratory (ORNL). N.Q.L. acknowledges support from the U.S. Naval Laboratory (NRL) through the National Research Council Research Associateship Programs. C.A.P. acknowledges support from the Laboratory Directed Research and Development Program of Oak Ridge National Laboratory, managed by UT-Battelle, LLC, for the U.S. Department of Energy. Computational work was performed using resources of the Advanced Research Computing Services at the University of Virginia and the ``Campus Compute Co-operative (CCC)"~\cite{Grimshaw2016}. The authors are grateful for useful discussions with Prof. Keivan Esfarjani, and LeighAnn Larkin.\\
R.R. and J.Z. contributed equally to this work.

\end{acknowledgments}

\appendix

\section*{Appendix A: Simulation Details}
We calculate thermal conductance between the left and right contacts of abrupt and mass-graded interfaces shown in Fig. \ref{fig1}. For each system, the crystal structure (FCC with a single atom per primitive unit cell), lattice constant $a$, and interatomic force constants are invariant. The boundaries between adjacent materials are perfectly abrupt. The junction atomic mass $m_j$ is varied between the contact atomic masses $m_l$ and $m_r$ according to Eq.~\ref{eqnexp}.

To describe atomic interactions, we use the Lennard-Jones (LJ) potential $U_{LJ}(r_{ij}) = 4\epsilon[(\sigma/r_{ij} )^{12} - (\sigma/r_{ij} )^6]$, where $r_{ij}$ is the distance between atoms i and j, $\epsilon = 0.0503$ eV is the energy scale, and $\sigma = 3.37 \AA$ is the length scale. These potential parameters are chosen to be identical to those used in our previous work~\cite{Polanco2017}. The cutoff distance for the LJ potential is chosen to be 2.5$\sigma$ including interactions up to $5^{th}$ nearest neighbors. In the harmonic NEGF calculations, we use $2^{nd}$ order expansion of this potential to model atomic interactions. The equilibrium lattice constant for this structure is a = 5.22 $\AA$ at \textbf{T} = 0 K. The mass of the left contact is fixed to $m_l = 40$ amu, while the mass of the right contact is either 120 amu (Figs. 2-8) or 400 amu. (Fig. 9).

For the NEMD simulations, we use the LAMMPS MD simulator on a system with $10\times10\times302$ conventional unit cells and a time-step of 2 fs. We impose periodic boundary conditions over x and y directions and set the atomic layers at the two ends of the system as walls. Heat is added to the system from the left edge and removed from the right edge using the Langevin thermostat. The baths' temperatures are set to $\textbf{T}_{bath} = (1\pm0.1)\textbf{T}$ with a time constant of 1.07 ps over blocks of 50 unit cells thick. This setup for the thermostat is done to ensure sufficient phonon-phonon scattering that prevents potential size effects especially at low temperatures. On the computations at very low temperatures (\textbf{T} = 2 K), significant size effects can arise due to the lack of phonon thermalization. \textbf{T} = 2 K corresponds to the non-dimensional temperature $k_B \textbf{T}/\epsilon = 0.003$, which is less than 1\% of the melting temperature. At such a low temperature, atomic displacements are small and atoms behave almost harmonically. Thus, phonon interactions due to anharmonicity are small and phonon thermalization is small. To obtain enough thermalization in our system, we increase the size of each thermal bath. We tested for potential size effects due to the cross section, the thickness of the domain and the thermostat time constant in our previous work~\cite{Polanco2017} and no significant change in the interfacial thermal conductance was noticed. We also test for size effects on mass-graded interfaces. Table \ref{sizeeffect} exhibits how thermal conductance changes with the thickness of the simulation domain increasing from 60 to 300 unit cells. It is indicated in Table \ref{sizeeffect} that the change in $G$ is less than the standard deviation of our NEMD simulations results. Thus we conclude that the choice of 300 unit cells for our simulation domain thickness guarantees that our results are size effect independent.

\begin{table}[h!]
\caption{Size effects on the thermal conductance of a mass-graded interface determined by our NEMD simulations. Results are for the system with $m_l = 40$ amu, $m_r = 120$ amu and the mass-graded layer with 10 layers $N_l$ and each layer thickness $t$ is equal to 2 unit cells. The conductance values are given in MW $m^{-2}$ $K^{-1}$. The standard deviation of five independent calculations is the reported error.}
\centering
\begin{adjustbox}{width=1\columnwidth}
\small
\begin{tabular}{ l c c c c c }
 \hline
 \hline
 \\
  & & & $G$ (MW $m^{-2}$ $K^{-1}$) & & \\
 \\
 \hline
 \\
 Thickness (u.c.) & 60 & 90 & 120 & 240 & 300 \\
 \\
 \hline
 \\
 \textbf{T} = 2 K & 88.17 $\pm$ 0.47 & 88.86 $\pm$ 1.59 & 89.04 $\pm$ 0.59 & 91.06 $\pm$ 0.93 & 91.26 $\pm$ 0.65 \\
 \\
 \textbf{T} = 30 K & 146.47 $\pm$ 1.02 & 150.66 $\pm$ 1.09 & 155.01 $\pm$ 2.51 & 165.77 $\pm$ 5.45 & 163.95 $\pm$ 5.04 \\ 
 \\
 \hline
 \hline
\end{tabular}
\end{adjustbox}
\label{sizeeffect}
\end{table}

To prevent changes of pressure as the temperature varies from affecting thermal transport at the interface, we account for the thermal expansion of the system. We perform equilibration runs under zero pressure at different temperatures using the isothermal-isobaric ensemble (NPT). The results are used to find the dependence of the lattice constant with temperature, which is fitted to a third order polynomial function:
\begin{equation}
    a(\textbf{T}) = 5.2222 + 0.0004\textbf{T} +10^{-6} \textbf{T}^2 - 4\times10^{-9} \textbf{T}^3 \AA.
\end{equation}
Atoms are first equilibrated under the microcanonical ensemble (NVE) for 4 ns. Next, heat is added to the system for 10 ns to achieve steady state. Then, the temperature is recorded for 6 ns to ensure a proper statistical average. From the temperature profile, we estimate the interfacial thermal conductance dividing the heat flux over the temperature drop, which arises from a linear fit of the temperature at each lead extrapolated to the interface.\\
Conductance values from NEMD reported in this paper are averages over five independent calculations for which the initial velocity condition are generated randomly.

In the NEGF simulations,  we use 100 grid points to sample the 0--20 Trad/s frequency range and 100$\times$100 grid points to sample a FCC conventional unit cell Brillouin zone. This dense frequency and wavevector samplings guarantees the convergence of $G$. In all  NEGF calculations, we excluded the effect from the contact resistance, and the conductances are four-probe measured conductances~\cite{Polanco2017}. This set-up  helps us compare the NEGF and NEMD results. Before we started the systematic calculations of the exponential graded systems, we compared our NEGF and MD simulations. For an $m_l=40$ amu and $m_r=120$ amu abrupt interface system, we got $G=$59.3 MW m$^{-2}$ K$^{-1}$ at $T=$0 K from NEGF calculations and $G=$61.0 MW m$^{-2}$ K$^{-1}$ at $T=$2 K from NEMD. These values are reasonably close considering the temperature difference. 

\section*{Appendix B: Figures supporting the manuscript}

\begin{figure}[ht]
	\centering
	\includegraphics[width=76mm]{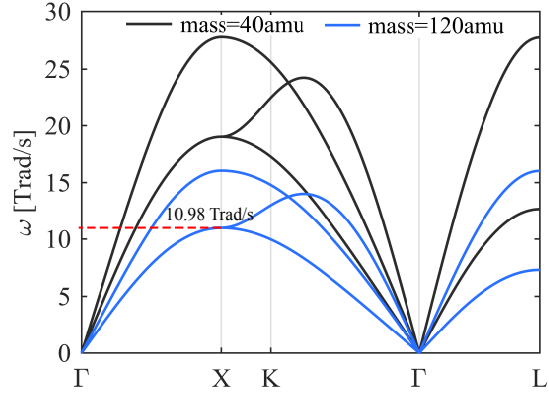}
	\caption{The range for phonon transport in the same polarization will be in energy range: 0--10.98 Trad/s for 2 TA bands in the left contact to 2 TA bands in the right contact; 0--16.06 Trad/s for 1 LA band to 1 LA band.  }
	\label{fig1S}
\end{figure}

\begin{figure}[ht]
	\centering
	\includegraphics[width=76mm]{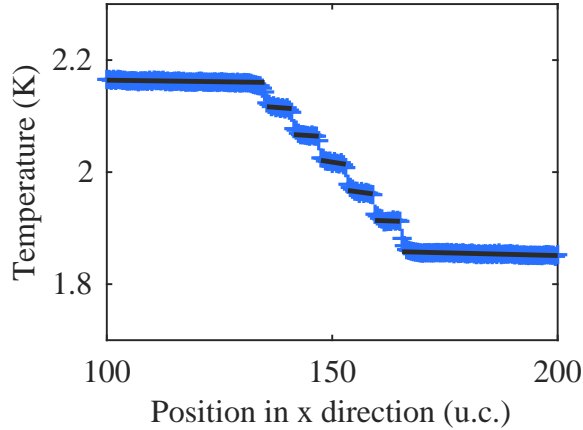}
	\caption{Temperature profile for $t$ = 6 u.c. and $N_l$ = 5. The temperature drop at interfaces is 93\% of the total drop between the contacts.}
	\label{fig2S}
\end{figure}

\section*{REFERENCES}
\bibliographystyle{ieeetr}
\bibliography{references_ver10}

\begin{thebibliography}{10}

\bibitem{Pop2010}
E.~Pop, ``{Energy Dissipation and Transport in Nanoscale Devices},'' {\em Nano
  Res}, vol.~3, pp.~147--169, 2010.

\bibitem{Riedel2009}
G.~Riedel, J.~Pomeroy, K.~Hilton, J.~Maclean, D.~Wallis, M.~Uren, T.~Martin,
  U.~Forsberg, A.~Lundskog, A.~Kakanakova-Georgieva, G.~Pozina, E.~Janzen,
  R.~Lossy, R.~Pazirandeh, F.~Brunner, J.~Wurfl, and M.~Kuball, ``{Reducing
  Thermal Resistance of AlGaN/GaN Electronic Devices Using Novel Nucleation
  Layers},'' {\em IEEE Electron Device Letters}, vol.~30, pp.~103--106, feb
  2009.

\bibitem{Pomeroy2014}
J.~W. Pomeroy, M.~Bernardoni, D.~C. Dumka, D.~M. Fanning, and M.~Kuball, ``{Low
  thermal resistance GaN-on-diamond transistors characterized by three-
  dimensional Raman thermography mapping},'' {\em Applied Physics Letters},
  vol.~104, p.~083513, 2014.

\bibitem{Sun2016}
H.~Sun, J.~W. Pomeroy, R.~B. Simon, D.~Francis, F.~Faili, D.~J. Twitchen, and
  M.~Kuball, ``{Temperature-Dependent Thermal Resistance of GaN-on-Diamond HEMT
  Wafers},'' {\em IEEE Electron Device Letters}, vol.~37, no.~5, pp.~621--624,
  2016.

\bibitem{Zhou2017}
Y.~Zhou, R.~Ramaneti, J.~Anaya, S.~Korneychuk, J.~Derluyn, H.~Sun, J.~Pomeroy,
  J.~Verbeeck, K.~Haenen, and M.~Kuball, ``{Thermal characterization of
  polycrystalline diamond thin film heat spreaders grown on GaN HEMTs},'' {\em
  Appl. Phys. Lett}, vol.~111, 2017.

\bibitem{Cho2017}
J.~Cho, D.~Francis, D.~H. Altman, M.~Asheghi, and K.~E. Goodson, ``{Phonon
  conduction in GaN-diamond composite substrates},'' {\em Journal of Applied
  Physics}, vol.~121, no.~055105, 2017.

\bibitem{Cahill2003}
D.~G. Cahill, W.~K. Ford, K.~E. Goodson, G.~D. Mahan, A.~Majumdar, H.~J. Maris,
  R.~Merlin, and S.~R. Phillpot, ``{Nanoscale thermal transport},'' {\em
  Journal of Applied Physics}, vol.~93, no.~793, 2003.

\bibitem{Hopkins2013}
P.~E. Hopkins, ``{Thermal Transport across Solid Interfaces with Nanoscale
  Imperfections: Effects of Roughness, Disorder, Dislocations, and Bonding on
  Thermal Boundary Conductance},'' {\em ISRN Mechanical Engineering},
  vol.~2013, pp.~1--19, jan 2013.

\bibitem{Cahill2014}
D.~G. Cahill, P.~V. Braun, G.~Chen, D.~R. Clarke, S.~Fan, K.~E. Goodson,
  P.~Keblinski, W.~P. King, G.~D. Mahan, A.~Majumdar, H.~J. Maris, S.~R.
  Phillpot, E.~Pop, and L.~Shi, ``{Nanoscale thermal transport. II.
  2003–2012},'' {\em Applied Physics Reviews}, vol.~1, 2014.

\bibitem{Merabia2014}
S.~Merabia and K.~Termentzidis, ``{Thermal boundary conductance across rough
  interfaces probed by molecular dynamics},'' {\em Physical Review B}, vol.~89,
  p.~54309, feb 2014.

\bibitem{Monachon2016}
C.~Monachon, L.~Weber, and C.~Dames, ``{Thermal Boundary Conductance: A
  Materials Science Perspective},'' {\em Annu. Rev. Mater. Res}, vol.~46,
  pp.~433--63, 2016.

\bibitem{Zhang2018}
J.~Zhang, C.~A. Polanco, and A.~W. Ghosh, ``{Optimizing the Interfacial Thermal
  Conductance at Gold-Alkane Junctions From “First Principles”},'' {\em
  Journal of Heat Transfer}, vol.~140, no.~9, p.~092405, 2018.

\bibitem{Rastgar2016}
R.~Rastgarkafshgarkolaei, Y.~Zeng, and J.~Khodadadi, ``{A molecular dynamics
  study of the effect of thermal boundary conductance on thermal transport of
  ideal crystal of n-alkanes with different number of carbon atoms},'' {\em
  Journal of Applied Physics}, vol.~119, no.~20, p.~205107, 2016.

\bibitem{Aryana2018}
K.~Aryana and M.~B. Zanjani, ``{Diamond family of colloidal supercrystals as
  phononic metamaterials},'' {\em Journal of Applied Physics}, vol.~123,
  no.~18, 2018.

\bibitem{Losego2012}
M.~D. Losego, M.~E. Grady, N.~R. Sottos, D.~G. Cahill, and P.~V. Braun,
  ``{Effects of chemical bonding on heat transport across interfaces},'' {\em
  Nature Materials}, vol.~11, 2012.

\bibitem{Hohensee2015}
G.~T. Hohensee, R.~B. Wilson, and D.~G. Cahill, ``{Thermal conductance of
  metal–diamond interfaces at high pressure},'' {\em Nature Communications},
  vol.~6, p.~6578, 2015.

\bibitem{Schmidt2010}
A.~J. Schmidt, K.~C. Collins, A.~J. Minnich, and G.~Chen, ``{Thermal
  conductance and phonon transmissivity of metal-graphite interfaces},'' {\em
  Journal of Applied Physics}, vol.~107, no.~10, 2010.

\bibitem{Saltonstall2013}
C.~B. Saltonstall, C.~A. Polanco, J.~C. Duda, A.~W. Ghosh, P.~M. Norris, and
  P.~E. Hopkins, ``{Effect of interface adhesion and impurity mass on phonon
  transport at atomic junctions},'' {\em Journal of Applied Physics}, vol.~113,
  p.~13516, 2013.

\bibitem{Duda2013}
J.~C. Duda, C.-Y.~P. Yang, B.~M. Foley, R.~Cheaito, D.~L. Medlin, R.~E. Jones,
  and P.~E. Hopkins, ``{Influence of interfacial properties on thermal
  transport at gold:silicon contacts},'' {\em Applied Physics Letters},
  vol.~102, p.~081902, 2013.

\bibitem{Jeong2016}
M.~Jeong, J.~P. Freedman, H.~J. Liang, C.-M. Chow, V.~M. Sokalski, J.~A. Bain,
  and J.~A. Malen, ``{Enhancement of Thermal Conductance at Metal-Dielectric
  Interfaces Using Subnanometer Metal Adhesion Layers},'' {\em Physical Review
  Applied}, vol.~5, p.~014009, jan 2016.

\bibitem{Lee2016a}
E.~Lee, T.~Zhang, T.~Yoo, Z.~Guo, and T.~Luo, ``{Nanostructures Significantly
  Enhance Thermal Transport across Solid Interfaces},'' {\em ACS Applied
  Materials and Interfaces}, vol.~8, pp.~35505--35512, dec 2016.

\bibitem{English2012b}
T.~S. English, J.~C. Duda, J.~L. Smoyer, D.~a. Jordan, P.~M. Norris, and L.~V.
  Zhigilei, ``{Enhancing and tuning phonon transport at vibrationally
  mismatched solid-solid interfaces},'' {\em Physical Review B - Condensed
  Matter and Materials Physics}, vol.~85, no.~3, pp.~1--14, 2012.

\bibitem{Tian2012}
Z.~Tian, K.~Esfarjani, and G.~Chen, ``{Enhancing phonon transmission across a
  Si/Ge interface by atomic roughness: First-principles study with the Green's
  function method},'' {\em Physical Review B - Condensed Matter and Materials
  Physics}, vol.~86, no.~23, pp.~1--7, 2012.

\bibitem{Polanco2015}
C.~A. Polanco, R.~Rastgarkafshgarkolaei, J.~Zhang, N.~Q. Le, P.~M. Norris,
  P.~E. Hopkins, and A.~W. Ghosh, ``{Role of crystal structure and junction
  morphology on interface thermal conductance},'' {\em Physical Review B},
  vol.~92, no.~144302, 2015.

\bibitem{saaskilahti2014}
K.~Saaskilahti, J.~Oksanen, J.~Tulkki, and S.~Volz, ``{Role of anharmonic
  phonon scattering in the spectrally decomposed thermal conductance at planar
  interfaces},'' {\em Physical Review B - Condensed Matter and Materials
  Physics}, vol.~90, no.~13, p.~134312, 2014.

\bibitem{Liang2011}
Z.~Liang and H.~Tsai, ``{Effect of thin film confined between two dissimilar
  solids on interfacial thermal resistance},'' {\em J. Phys.: Condens. Matter},
  vol.~23, 2011.

\bibitem{Smoyer2015}
J.~Smoyer, ``{Local Modification to Phononic Properties at Solid-Solid
  Interfaces : Effects on Thermal Transport},'' 2015.

\bibitem{Polanco2016}
C.~A. Polanco and A.~W. Ghosh, ``{Enhancing phonon flow through one-dimensional
  interfaces by impedance matching},'' vol.~083503, no.~2014, pp.~0--7, 2016.

\bibitem{Polanco2017}
C.~A. Polanco, R.~Rastgarkafshgarkolaei, J.~Zhang, N.~Q. Le, P.~M. Norris, and
  A.~W. Ghosh, ``{Design rules for interfacial thermal conductance: Building
  better bridges},'' {\em Physical Review B}, vol.~95, 2017.

\bibitem{lee2017a}
E.~Lee, T.~Yoo, and T.~Luo, ``{The role of intermediate layers in thermal
  transport across GaN/SiC interfaces},'' in {\em IEEE}, pp.~1--5, 2017.

\bibitem{Raut2011}
H.~K. Raut, V.~A. Ganesh, A.~S. Nair, and S.~Ramakrishna, ``{Anti-reflective
  coatings: A critical, in-depth review},'' {\em Energy {\&} Environmental
  Science}, vol.~4, no.~10, pp.~3779--3804, 2011.

\bibitem{Le2017}
N.~Q. Le, C.~A. Polanco, R.~Rastgarkafshgarkolaei, J.~Zhang, A.~W. Ghosh, and
  P.~M. Norris, ``{Effects of bulk and interfacial anharmonicity on thermal
  conductance at solid/solid interfaces},'' {\em Physical Review B}, vol.~95,
  no.~24, 2017.

\bibitem{Datta2005}
S.~Datta, {\em {Quantum Transport: Atom to Transistor}}.
\newblock Cambridge university press, 2005.

\bibitem{Clapham1973}
P.~B. Clapham and M.~C. Hutley, ``{Reduction of Lens Reflexion by the "Moth
  Eye" Principle},'' {\em Nature}, vol.~244, p.~281, 1973.

\bibitem{Bernhard1967}
C.~Bernhard, ``{Structural and functional adaptation in a visual system},''
  {\em Endeavour}, vol.~26, no.~98, pp.~79--84, 1967.

\bibitem{Zhou2016}
Y.~Zhou, X.~Zhang, and M.~Hu, ``{An excellent candidate for largely reducing
  interfacial thermal resistance: a nano-confined mass graded interface},''
  {\em Nanoscale}, vol.~8, no.~4, pp.~1994--2002, 2016.

\bibitem{Jeong2012}
C.~Jeong, S.~Datta, and M.~Lundstrom, ``{Thermal conductivity of bulk and
  thin-film silicon: A Landauer approach},'' {\em Journal of Applied Physics},
  vol.~111, p.~93708, 2012.

\bibitem{ghosh2016}
A.~Ghosh, {\em {Nanoelectronics: A Molecular View}}.
\newblock World Scientific Publishing Company, 2016.

\bibitem{Mingo2003}
N.~Mingo and L.~Yang, ``{Phonon transport in nanowires coated with an amorphous
  material: An atomistic Green's function approach},'' {\em Physical Review B},
  vol.~68, no.~24, p.~245406, 2003.

\bibitem{Wang2008}
J.-S. Wang, J.~Wang, and J.~T. L{\"{u}}, ``{Quantum thermal transport in
  nanostructures},'' {\em The European Physical Journal B}, vol.~62, no.~4,
  pp.~381--404, 2008.

\bibitem{Polanco2013}
C.~A. Polanco, C.~B. Saltonstall, P.~M. Norris, P.~E. Hopkins, and A.~W. Ghosh,
  ``{Impedance Matching of Atomic Thermal Interfaces Using Primitive Block
  Decomposition},'' {\em Nanoscale and Microscale Thermophysical Engineering},
  vol.~17, no.~3, pp.~263--279, 2013.

\bibitem{Wu1994}
C.~L. Wu, W.~C. Hsu, H.~M. Shieh, and W.~C. Liu, ``{Mobility enhancement in
  double $\delta$-doped GaAs/In x Ga 1-x As/GaAs pseudomorphic structures by
  grading the heterointerfaces},'' {\em Appl. Phys. Lett}, vol.~64, no.~3027,
  1994.

\bibitem{HoiWong2014}
M.~{Hoi Wong}, S.~Keller, H.~Li, M.~Laurent, Y.~Hu, N.~Pfaff, J.~Lu, D.~F.
  Brown, N.~A. Fichtenbaum, J.~S. Speck, S.~P. DenBaars, and U.~K. Mishra,
  ``{Recent progress in metal-organic chemical vapor deposition of (0001)
  N-polar group-III nitrides},'' {\em Semiconductor Science and Technology},
  vol.~29, no.~113001, 2014.

\bibitem{Grimshaw2016}
A.~Grimshaw, M.~A. Prodhan, A.~Thomas, C.~Stewart, and R.~Knepper, ``{Campus
  Compute Co-operative (CCC): A service oriented cloud federation},'' in {\em
  2016 IEEE 12th International Conference on e-Science (e-Science)}, pp.~1--10,
  IEEE, oct 2016.

\end{thebibliography}
\end{document}